\documentclass[12pt]{article}
\usepackage{graphicx}
\def\beq{\begin{equation}}
\def\eeq{\end{equation}}
\def\beqa{\begin{eqnarray}}
\def\eeqa{\end{eqnarray}}
\def\beq{\begin{equation}}
\def\eeq{\end{equation}}
\def\beqa{\begin{eqnarray}}
\def\eeqa{\end{eqnarray}}
\newcommand{\be}{\begin{equation}}
\newcommand{\eq}{\end{equation}}

\newcommand{\ee}{ \end{equation}}
\newcommand{\ba}{\begin{array}}
\newcommand{\ea}{\end{array}}

\def\R {\bf{R}}

\begin{document}
\begin{flushright}
CAMS-98/03\\
HUB-EP-98/44 \\
QMW-PH-98-33\\
hep-th/9807187   
\end{flushright}
\vspace{1cm}
\begin{center}
\baselineskip=16pt
\centerline{\bf BPS BLACK HOLES IN N=2 FIVE DIMENSIONAL }
\centerline{\bf  ADS SUPERGRAVITY}
\vspace{2truecm}
\centerline{\bf
K. Behrndt$^a$\footnote{behrndt@qft2.physik.hu-berlin.de}\ , \ A.
H. Chamseddine$^{b,c}$\footnote{chams@layla.aub.edu.lb}\
\
 and  \
W. A. Sabra$^{b,d}$\footnote{w.sabra@qmw.ac.uk}}
\vspace{.5truecm}
{\em
\centerline{$^a$Humboldt-Universit\"at, Institut f\"ur Physik}
\centerline{Invalidenstra\ss e 110, 10115 Berlin, Germany}

\centerline{$^b$Center for Advanced Mathematical Sciences, 
American University of
 Beirut, Lebanon.$$\footnote{
Permanent address}}

\centerline{$^c$ Institute for Theoretical Physics, ETH Zuerich Switzerland.}
\centerline{$^d$Physics Department, Queen Mary and Westfield College,}
\centerline{Mile End Road, E1 4NS, London, United Kingdom}}
\end{center}

\vskip 1 cm
\begin{abstract}
BPS black hole solutions of $U(1)$ gauged five-dimensional
supergravity are obtained
by solving the Killing spinor equations.
These extremal static black holes live in an
asymptotic $AdS_5$ space time. Unlike black holes in asymptotic flat
space time none of them possess a regular horizon.
We also calculate the influence, of a particular class of these solutions,
on the Wilson loops calculation.
\end{abstract}
\bigskip

\bigskip
\newpage

In the past years a considerable amount of work has been devoted to
establish a duality between supergravity and super Yang Mills
theories. For example the conformal field theory ($CFT$) living on the
boundary of the five-dimensional anti-de Sitter space ( $AdS_5$) is
expected to be dual (in certain limits) to the four-dimensional super
 Yang Mills theory. Since this conjecture has been made \cite{010} and
further developed in \cite{030}, five-dimensional anti-de Sitter spaces
have received a great deal of interest.

The aim of this letter is to describe BPS black holes living in an
asymptotic $AdS_5$ vacuum (for the $AdS_4$ case, Reissner-Nordstr{\" o}m
solutions have been discussed in \cite{031} and non-abelian monopoles
in \cite{032}). To keep these solutions as general as
possible we formulate them in terms of $D$=5, $N$=2 supergravity with
an arbitrary prepotential, i.e.\ the $N=4, 8$ black holes appear as a
special subclass for special choices of the prepotential of the $N=2$
theory.  Because the asymptotic vacuum should be $AdS_5$ instead of
flat Minkowski space, we gauge a $U(1)$ subgroup of the $SU(2)$
automorphism group, which results in a scalar potential that becomes
constant at infinity.

In the first part we will describe these black holes as solutions of the
Killing spinor equations of gauged $D$=5, $N$=2 supergravity \cite{cn} 
and in the
second part we ask for the modification of the Wilson-loop calculations as
done in \cite{rey}, \cite{malda}, \cite{020}.

First, we briefly describe the theory of $N=2$ supergravity coupled to
an arbitrary number $n$ of abelian supermultiplets. $N=2$ supergravity
theories in five-dimensions can be obtained, for example, by
compactifying eleven-dimensional supergravity on a Calabi-Yau 3-folds
\cite{Cadavid}. The massless spectrum of the compactified theory
contains $(h_{(1,1)}-1)$ vector multiplets with real scalar
components. Including the graviphoton, the theory has $h_{(1,1)}$
vector bosons. The theory also contains $h_{(2,1)}+1$ hypermultiplets,
where $h_{(1,1)}$ and $h_{(2,1)},$ are the Calabi-Yau Hodge
numbers. In what follows and for our purposes the hypermultiplets are
switched off. The anti-de  Sitter supergravity  can be obtained by
gauging the $U(1)$ subgroup of the $SU(2)$ automorphism group of the
$N=2$ supersymmetry algebra. This gauging, which breaks $SU(2)$ down
to $U(1)$ can be achieved by introducing a linear combination of the
abelian vector fields present in the ungauged theory,
i.e. $A_\mu=V_IA_\mu^I$, with a coupling constant $g$. To restore
supersymmetry, $g$-dependent and gauge-invariant terms have to be
added. In a bosonic background, this amounts to the addition of a
scalar potential, (for more details see \cite{GST,TA}).

The bosonic part of the effective gauged supersymmetric $N=2$ Lagrangian
which describes the coupling of vector multiplets to supergravity is given
by
\begin{eqnarray}
e^{-1} \mathcal{L} &=& -{\frac{1}{2}} R - {\frac{1}{4}} G_{IJ} F_{\mu\nu}
{}^I F^{\mu\nu J}-{\frac{1}{2}} g_{ij} \partial_{\mu} \phi^i \partial^\mu
\phi^j +{\frac{e^{-1}}{48}} \epsilon^{\mu\nu\rho\sigma\lambda} C_{IJK}
F_{\mu\nu}^IF_{\rho\sigma}^JA_\lambda^k  \nonumber \\
&+&   g^2V_IV_J\Big(6X^IX^J - {9 \over 2}\,
g^{ij}\partial_iX^I\partial_jX^J\Big)
\label{action}
\end{eqnarray}
where $R$ is the scalar curvature, $F_{\mu\nu}=2\partial_{[\mu}A_{\nu]}$ is
the Maxwell field-strength tensor and $e=\sqrt{-g}$ is the determinant of
the F\"unfbein $e_m^{\ a}$. \footnote{
The signature $(-++++)$ is used. Antisymmetrized indices are defined by: 
$[ab] = {\frac{1 }{2}} (ab -ba)$.}

The physical quantities in (\ref{action}) can all be expressed in terms of a
homogeneous cubic polynomial ${V}$ which defines \lq\lq very special
geometry'' \cite{very}.
\begin{equation}
G_{IJ} = -{\frac{1}{2}}{\frac{\partial}{\partial X^I}} {\frac{\partial}{%
\partial X^J}}(\ln {V})|_{{V} =1},\qquad g_{ij} = G_{IJ}
\partial_{i}X^I\partial_{j}X^J|_{{V} =1},\qquad (\partial_i \equiv {\frac{%
\partial }{\partial\phi^i}}).  \label{metric}
\end{equation}
For Calabi-Yau compactification
\begin{equation}
{V} = {\frac{1}{6}} C_{IJK} X^I X^J X^K=X^IX_I=1.  \label{pin}
\end{equation}
${V}$ is the intersection form, $X^I$ and $X_I$ correspond to the size of
the 2 and 4-cycles and $C_{IJK}$ are the intersection numbers of the
Calabi-Yau threefold.

Since we are interested in finding BPS solutions in the gauged theory, we
display the supersymmetry transformation of the Fermi fields in a bosonic
background
\begin{eqnarray}
\delta\psi_\mu &=&\Big(\mathcal{D}_\mu + {\frac{ i}{8}}X_I (\Gamma_\mu{}%
^{\nu\rho} - 4 \delta_\mu{}^ \nu \Gamma^\rho) F_{\nu\rho}{}^I+{\frac{1}{2}}%
g\Gamma_\mu X^IV_I-{\frac{3}{2}}igV_IA^I_\mu\Big) \epsilon,  \nonumber \\
\delta \lambda _i& =&\Big({\frac{3}{8}}\partial_i X_I\Gamma^{\mu\nu}
F_{\mu\nu}^I - {\frac{i}{2}} g_{ij} \Gamma^\mu \partial_\mu\phi^j+
{\frac{3}{2%
}}igV_I \partial_iX^I\Big)\epsilon
\end{eqnarray}
where $\epsilon$ is the supersymmetry parameter and $\mathcal{D}_\mu$ is the
covariant derivative

The spherically symmetric BPS electric solutions can be obtained by solving
for the vanishing of the gravitino and gaugino supersymmetry variation for a
particular choice for the supersymmetry parameter. We impose the projection
operator condition on the spinor $\epsilon $
\begin{equation} \label{projector}
\epsilon =\Big(ia\Gamma _{0}+b\Gamma _{1}\Big)\epsilon ,
\end{equation}
where $a^{2}+b^{2}=1$ and this  breaks $N=2$ supersymmetry to $N=1$.

We briefly\footnote{More detailed analysis will be given in \cite{TA}.}
 describe the  procedure of obtaining solutions preserving $N=1$
supersymmetry. First we start with an ansatz for the metric and gauge field
\[
ds^{2}=-e^{2V}dt^{2}+e^{2W}\left( dr^{2}+f^{2}r^{2}\left( d\theta ^{2}+\sin
^{2}\theta d\phi ^{2}+\cos ^{2}\theta d\psi ^{2}\right) \right)
\]
\[
A_{t}^{I}=e^{-2U}X^{I}
\]
where the functions $U,V,W$ and $f$ are functions of $r$, and
$(\theta ,\phi ,\psi )$ are the polar coordinates of the 3-sphere. 
 As solution of the gauge field equations we find
\[
e^{2U}X_{I}=\frac{1}{3}H_{I}
\]
where $H_{I}$ is a harmonic functions which depends on the electric
charge $q_I$ . The supersymmetry variation of the gaugino and the time
component of the gravitino imply the following relations
\[
e^{2V}=e^{-4U}f^{2}\ ,\ \qquad e^{2W}=e^{2U}\frac{1}{f^{2}}
\]
\[
a=-\frac{1}{f}\ , \qquad b=-\frac{1}{f}gre^{3U}\ , \qquad
f^2 = 1 + g^2 r^2 e^{6U}
\]
where we used some relations of special geometry analog to the
derivation in \cite{040}. 

The time and spatial components of the gravitino transformation imply 
differential constraints on the Killing spinor. These are
\[
\left( \partial _{t}-ig\right) \epsilon =0 \ ,
\]
\[
\left( \partial _{r}-\frac{i}{2f}\left( \frac{1}{r}+3U^{\prime }\right)\Gamma_0 -%
\frac{1}{2}\left( \frac{1}{r}+U^{\prime }\right)\right) \epsilon
=0,
\]
\[
\left( \partial _{\theta }+\frac{i}{2}\Gamma _{012}\right) \epsilon =0 \ ,
\]
\[
\left( \partial _{\phi }+\frac{i}{2}\sin \theta \Gamma _{013}-\frac{1}{2}%
\cos \theta \Gamma _{23}\right) \epsilon =0,
\]
\[
\left(   \partial_\psi   +\frac{i}{2}\cos \theta 
\Gamma _{014}+\frac{1}{2}\sin
\theta \Gamma _{24}\right) \epsilon =0.
\] 
Going to the rescaled coordinates
\[
X^{I}=\mathcal{V}^{-{\frac{1}{3}}}Y^{I}
\]
where $\mathcal{V}=e^{3U}$, one obtains the following solution\footnote{The
Killing spinors for a general $AdS_p \times S^q$ geometry are also
discussed in \cite{042}.}
\begin{eqnarray}
\epsilon  &=&e^{igt}e^{-{\frac{i}{2}}\Gamma _{012}\theta }e^{{\frac{1}{2}}%
\Gamma _{23}\phi }e^{-{\frac{i}{2}}\Gamma _{014}\psi }\varphi (r) 
\nonumber\\
\varphi (r) &=&{1\over2}{{\cal V}^{-{1\over2}}\over\sqrt{gr}}
\Big(\sqrt{f+1}-\sqrt{f-1}\Gamma_1\Big) e^{{1\over2}\int^{r}
d\bar r \Big({1\over \bar r}+{1\over 3}{{\cal V}'\over{\cal V}}  \Big)} 
\; (1-i\Gamma_0)\epsilon_0 \,
\end{eqnarray}
where $f=\sqrt{1+g^{2}r^{2}\mathcal{V}^{2}}$ and
$\epsilon_0$ is an arbitrary constant spinor. Thus, inserting all terms
in our ansatz one obtains
\begin{eqnarray}
ds^{2} &=&-\mathcal{V}^{-4/3}(1+g^{2}r^{2}\mathcal{V}^{2})dt^{2}+\mathcal{V}%
^{2/3}\left[ {\frac{dr^{2}}{1+g^{2}r^{2}\mathcal{V}^{2}}}+r^{2}(d\theta
^{2}+\sin ^{2}\theta d\phi ^{2}+\cos ^{2}\theta d\psi ^{2})\right]
\nonumber \\
F_{tm}^{I} &=&-\partial _{m}(\mathcal{V}^{-1}Y^{I}) \qquad , \qquad
 \nonumber \\
\mathcal{V} &=&{\frac{1}{6}}C_{IJK}Y^{I}Y^{J}Y^{K},\qquad {\frac{1}{2}}%
C_{IJK}Y^{J}Y^{K}=H_{I}=3V_{I}+{\frac{q_{I}}{r^{2}}}  \label{solution}
\end{eqnarray}
Note that the constant parts in the harmonic functions are given by
$V_I$, which fixes the $U(1)$ that has been gauged.  The only
deviation from the ungauged case \cite{050} comes via the function
$f^2 = 1 + g^2 r^2 {\cal V}^2$. This term however changes completely
the singularity structure of the black hole solution. To investigate
this in more detail we may   consider simple cases were
${\cal V}$ can be written as  
\begin{equation}
\mathcal{V} = H^n = \Big(1 + {\frac{q}{r^2}}\Big)^n \qquad , \qquad n =
0,1,2,3\ .
\end{equation}
Obviously, the first case ($n=0$) defines the $AdS_5$ vacuum with no
black hole. The cases of $n=1,2$ correspond to black holes with a
singular horizon and they appear naturally as BPS solutions of $N=4,
8$ supergravity. In both cases the scalars are either zero or blow up
near the horizon.   The last case ($n=3$) is an example of a BPS black
hole of $N=2$ supergravity, which   seems to have a regular horizon at
$r\simeq 0$. However this coordinate system is misleading. Defining
\begin{equation}
\rho^2 = r^2 + q
\end{equation}
one finds
\begin{equation}  \label{020}
\begin{array}{l}
ds^2 = - e^{2V} dt^2 + e^{-2V} \Delta^{-1} \, d\rho^2 + \Delta \, \rho^2
d\Omega_3 \\
e^{2V} = \tilde H^{\frac{2n }{3}} + g^2 \rho^2 \Delta \quad , \quad \Delta =
\tilde H^{\frac{3-n }{3}}\quad , \quad \tilde H = 1 - {\frac{q}{\rho^2}}
\end{array}
\end{equation}
In the ungauged case ($g$=0) the horizon is at $\tilde H =0$ (or $\rho^2
= q$), which is regular in the case $n$=3 or $\Delta=1$. But taking
into account the gauging   the horizon disappeared ($e^{\pm 2V}$ is
finite at $\rho^2=q$ for $n=3$) and the singularity at $\rho=0$ becomes
naked.  For the other cases ($n=1,2$) the horizon becomes singular. For
$n=1$ the singular horizon is infinitely far away, i.e.\ a light
signal ($ds^2 =0$) would need infinite time to reach any finite
distance (null singularity). But for $n=2$ the distance to the
singular horizon is finite. This is different to the ungauged case
($g=0$), where all singular  cases  have null horizons. Note also, the
naked singularity at $\rho=0$ for $n=3$ (i.e.\ $\Delta =1$) is only a
finite distance away! Certainly, this makes this solution rather
suspisious and to overcome this situation one should     consider
the non-extremal case.

\begin{figure}[tbp]
\begin{center}
\includegraphics[angle=0, width=70mm]{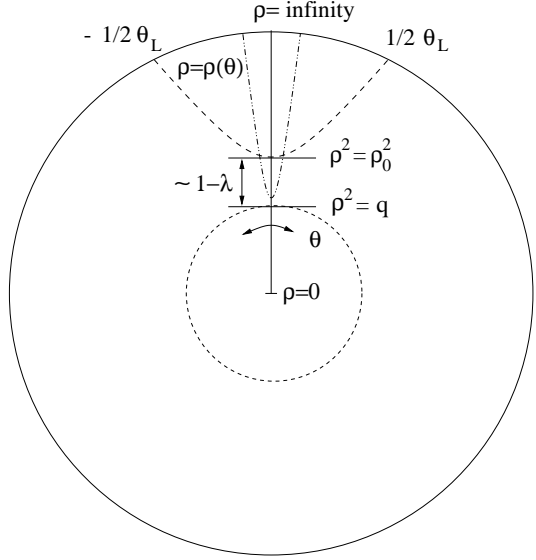}
\end{center}
\caption{This is our geometry: a open string attached with the endpoints at
asymptotic boundary, which is ``perturbed'' by a BPS black hole in the
middle. The dotted line should indicate the horizon. }
\end{figure}

Let us nevertheless ask, what is the influence of this black hole on the
Wilson loops as calculated in \cite{rey}, \cite{malda}, \cite{020}. For 
this we
calculate the Nambo-Goto action for open strings that are attached to the
asymptotic boundary
\begin{equation}
S = {\frac{1 }{2\pi \alpha^{\prime}}} \int d\tau d\sigma \sqrt{|\det
g_{\alpha\beta}|} \quad , \quad g_{\alpha\beta} = \partial_{\alpha} X
\partial_{\beta} X^N G_{MN}
\end{equation}
where $G_{MN}$ is the 5d metric. For the worldsheet coordinates we choose
the gauge
\begin{equation}
\tau = t \qquad , \qquad \sigma = \theta
\end{equation}
where $\theta$ is the polar angle in $\Omega_3$ (see figure and the
eq.  (\ref{solution})). Obviously, the string will stretch inside the
AdS space and thus its position is given by a function $f(\theta ,
\rho) =0$, where $\rho$ is the radial coordinate. This defining
equation can also be expressed as $\rho = \rho (\sigma = \theta)$. For
the induced metric we find therefore
\begin{equation}
\begin{array}{l}
g_{\tau\tau} = \partial_{\tau} X^M \partial_{\tau} X^N G_{MN} = G_{00} =
-e^{2V} \\
g_{\sigma\sigma} = \partial_{\sigma} \rho \partial_{\sigma} \rho\,
G_{\rho\rho} + G_{\theta\theta} = (\rho^{\prime})^2 e^{-2V}\Delta^{-1} +
\rho^2 \Delta
\end{array}
\end{equation}
and thus,
\begin{equation}
S = {\frac{1 }{2\pi \alpha^{\prime}}} \int d\sigma d\tau \sqrt{
(\rho^{\prime})^2 \, \Delta^{-1} + \rho^2 e^{2V} \Delta } \ .
\end{equation}
Following  arguments given by Maldacena \cite{malda}, we use the fact
that the Lagrangian does not depend explicitly on $\theta$ and therefore
\begin{equation}  \label{120}
c = {\frac{\rho^2 e^{2V} \Delta }{\sqrt{ (\rho^{\prime})^2 \Delta^{-1} +
\rho^2 e^{2V} \Delta}}} \ .
\end{equation}
The constant $c$ can be determined by going at the extremum $\rho_0$ where $%
\rho^{\prime}= 0$ i.e.
\begin{equation}
c^2 = \Big( \rho^2 \, e^{2V} \, \Delta \Big)_{\rho = \rho_0} \ .
\end{equation}
In addition it follows from (\ref{120}) that
\begin{equation}  \label{040}
d \sigma = {\frac{d\rho }{\rho \Delta e^{V} \sqrt{{\frac{1 }{c^2}} \rho^2
e^{2V} \Delta - 1 }}} = {\frac{dy }{2 g \, \rho_0 \, y^{\frac{n }{2}} (y -
\lambda)^{{\frac{3}{4}}\delta} \sqrt{y^{\frac{2n }{3}} \Big({\frac{y-\lambda
}{1 - \lambda}}\Big)^{\delta} -1}}}
\end{equation}
where $y = (\rho/ \rho_0)^2$, $\lambda = q / \rho_0^2$ and $\delta = {
\frac{2(3-n)}{3}}$ (see also the figure). Furthermore we consider here
only the simplest case where $e^{2V} = g^2 \rho^2 \Delta$ (i.e.\ we
neglect the first term), which is a good approximation for the region
$0 \ll q < \rho^2$. Integrating this equation yields the function
$\rho = \rho(\sigma = \theta)$ that determines the position of
the string in the AdS space. We can also calculate the distance
between both endpoints on the boundary\footnote{Note, we are dealing
here with a different asymptotic geometry of $\R \times S_3$ (where
$\R$ is the time).}
\begin{equation}
L = 2 \int_0^{\frac{\theta_L }{2}} \sqrt{g_{\sigma \sigma} d\sigma^2} = {%
\frac{(1 - \lambda)^{\frac{\delta }{2}} }{g}} \int_1^{\infty} {\frac{dy }{y^{%
\frac{n }{3}} (y - \lambda)^{\frac{\delta }{2}}\sqrt{ y^{\frac{2n }{3}}
(y-\lambda )^{\delta} - (1 - \lambda)^{\delta} }}} \ .
\end{equation}
There are some interesting things to notice. First, for $n=3$ ($\delta =0$)
all $\lambda$ dependence drops out and $L \sim 1/g$ becomes independent of $%
\rho_0$ and the charge $q$, it scales only with cosmological constant. Thus
it coincides with the case without black hole.
Secondly, for $\delta \neq 0$ the integral is finite if $\lambda < 1$, i.e.\
the string is away from the horizon. However, if the horizon comes close to
string ($\lambda \rightarrow 1$) the integral becomes divergent for $n=1$.
However taking the pre-factor into account one finds that in this limit $L$
behaves like $L \sim {\frac{1 }{g}}(1-\lambda)^{1 - {\frac{\delta }{2}}}$.
Therefore, the string endpoints approach each other $L \rightarrow 0$ for $q
\rightarrow \rho_0^2$ (see figure) if the horizon becomes large enough.
This is different from the (neutral) Schwarzschild black hole, where $L
\rightarrow \infty$ if the horizon comes close to the string \cite{020}.

Finally, one may insert the solution (\ref{040}) in the action and calculate
the energy
\begin{equation}
E ={\frac{ T }{2 \pi}} \rho_0^2 \int_1^{\infty} dy \left[ {\frac{y^{n/6} (y
- \lambda)^{\delta/4} }{\sqrt{ y^{2n/3} (y -\lambda)^{\delta} - (1 -
\lambda)^{\delta}}}} - {\frac{1 }{\sqrt{y}}} \right]
\end{equation}
where the last term is the subtraction of the infinite self-energy
(see \cite {malda}, \cite{020}). Obviously, for $\lambda \rightarrow
1$ or $q \rightarrow \rho_0^2$ the energy remains finite although the
string comes close to the singular horizon and it scales with the
charge or the BPS mass of the black hole $E \sim q$.  It is
interesting to note that the energy is independent of $g$.

\bigskip

{\bf Acknowledgements}

\medskip

We thank P.\ Townsend for bringing to our attention the
reference \cite{060}, which deals with black hole solutions
in (anti) de Sitter background without vector multiplets.



\begin{thebibliography}{9}
\bibitem{010}  J. Maldacena, \emph{The large N limit of superconformal field
theory and supergravity}, \texttt{hep-th/9711200}.

\bibitem{030}  S. S.\ Gubser, I. R.\ Klebanov and A. M. Polyakov, \emph{Gauge
theory correlators from non-critical string theory}, \texttt{hep-th/9802109}.%
\newline
E.\ Witten, \emph{Anti-de Sitter space and holography}, \texttt{%
hep-th/9802150}

\bibitem{031}
        L. J. Romans,
	\textit{Nucl.Phys.} \textbf{B383} (1992) 395,
        \texttt{hep-th/9203018}.

\bibitem{032}
        A. H. Chamseddine and M. S. Volkov,
	\textit{Phys.Rev.} \textbf{D57} (1998) 6242,
        \texttt{hep-th/9711181};
	\textit{Phys.Rev.Lett.} \textbf{79} (1997) 3343,
        \texttt{hep-th/9707176}.

\bibitem{cn} A. H. Chamseddine and H. Nicolai, 
	\textit{Phys. Lett.} \textbf{B96} (1980) 89 and unpublished
	notes.

\bibitem{rey}
	S.-J.\ Rey, 
	\emph{Macroscopic strings as heavy quarks in large N 
	gauge theory and anti-de Sitter supergravity}, 
	\texttt{hep-th/9803001}.

\bibitem{malda}  J. Maldacena, 
	\textit{Phys.Rev.Lett.} \textbf{80} (1998) 4859, 
	\texttt{hep-th/9803002}.

\bibitem{020}  S.-J. Rey, S. Theissen and J.-T. Yee, \texttt{hep-th/9803135}
\newline
A. Brandhuber, N. Itzhaki, J. Sonnenschein and S. Yankielowicz, \texttt{%
hep-th/9803138}, \texttt{hep-th/9803263}.

\bibitem{Cadavid}  A. C. Cadavid, A. Ceresole, R. D'Auria, and S. Ferrara,
\textit{Phys. Lett.} \textbf{B357} (1995) 76.

\bibitem{GST}  M. G\"{u}naydin, G. Sierra, and P. K. Townsend, \textit{Nucl.
Phys.} \textbf{B242} (1984) 244; \textit{Nucl. Phys.} \textbf{B253} (1985)
573.

\bibitem{TA}  K. Behrndt, A. H. Chamseddine and W. A. Sabra, to appear

\bibitem{very}  B. de Wit and A. Van Proyen, \textit{Phys. Lett.} \textbf{293%
} (1992) 94.


\bibitem{040} A. H. Chamseddine and W. A. Sabra, \textit{Phys. Lett.} 
	\textbf{B426} (1998) 36, \texttt{hep-th/9811161}.

\bibitem{042}
        H. Lu, C.N. Pope and J. Rahmfeld,
        \emph{A Construction of Killing spinors on $S^n$},
        \texttt{hep-th/9805151}.


\bibitem{050} W. A. Sabra,
\textit{Mod. Phys. Lett.} \textbf{A13} (1998) 239, \texttt{hep-th/9708103}

\bibitem{060}
	L.A.J. London,
	\emph{Arbitrary dimensional cosmological multi - black holes},
	\textit{Nucl. Phys.} \textbf{B434} (1995) 709.

\end{thebibliography}
\end{document}